\documentclass[9pt,twocolumn,twoside]{osajnl}
\usepackage{siunitx}
\DeclareSIUnit{\belmilliwatt}{Bm}
\DeclareSIUnit{\dBm}{\deci\belmilliwatt}
\usepackage{enumitem}
\usepackage{amsmath}

\journal{ol} 

\setboolean{shortarticle}{true}

\title{Photorefraction-induced Bragg scattering in cryogenic lithium niobate ring resonators}

\author[1]{Yuntao Xu}
\author[1]{Ayed Al Sayem}
\author[1]{Chang-Ling Zou}
\author[2]{Linran Fan}
\author[1,*]{Hong X. Tang}

\affil[1]{Department of Electrical Engineering, Yale University, New Haven, Connecticut 06520, USA}
\affil[2]{James C. Wyant College of Optical Sciences, The University of Arizona, Tucson, Arizona 85721, USA}

\affil[*]{Corresponding author: hong.tang@yale.edu}

\begin{abstract}
We report intracavity Bragg scattering induced by photorefractive (PR) effect in high-Q lithium niobate (LN) ring resonators at cryogenic temperatures. We show that, when a cavity mode is strongly excited, the PR effect imprints a long-lived periodic space charge field. This residual field in turn creates a refractive index modulation pattern that dramatically enhances the back scattering of an incoming probe light, and results in selective and reconfigurable mode splittings. This PR-induced Bragg scattering effect, despite being undesired for many applications, could be utilized to enable optically programmable photonic components.
\end{abstract}

\setboolean{displaycopyright}{true}

\begin{document}

\maketitle


Lithium niobate (LN), as one of the most widely studied optical materials, has played a significant role in nonlinear optics \cite{arizmendi2004photonic} for its rich and favorable optical properties \cite{weis1985lithium}.  One unique characteristics of LN is the photorefractive (PR) effect, which arises as a combination of the photo-excited space-charge field and the subsequent electro-optic effect, inducing a refractive index variation during light illumination \cite{weis1985lithium,hall1985photorefractive}. In the past decades, vast amount of research has been performed to study and control this important feature of LN crystals \cite{von1978intrinsic,volk1994optical,kong2012recent}. On one hand, the PR effect is considered to be responsible for the optical damage \cite{volk1994optical}, introducing instability and limiting the power handling capability of devices \cite{kong2012recent,lu2019periodically}. On the other hand, the photorefractive effect could also be utilized for optical holography and storage \cite{guo2004improvement,yariv1996holographic}.

With recent development of smart-cut wafer-bonding technique \cite{bazzan2015optical}, high-quality thin film single-crystalline LN on insulator (LNOI) has enabled on-chip high quality factor (Q) LN resonators \cite{zhang2017monolithic}, and provided a promising chip-scale system for various nonlinear optics applications \cite{gong2019soliton,wang2019monolithic,lu2019periodically,wang2017second,he2019self}. Due to the narrow linewidth of the resonances, LN high-Q microresonators also offer an opportunity to study the refractive index variation induced by the PR effect. The enhanced intracavity optical intensity significantly boosts the PR effect and leads to observations of novel phenomena such as photon-level tuning of resonances and quenching of the PR effect \cite{sun2017nonlinear,jiang2017fast,liang2017high,li2019photon}. 

\begin{figure}[!t]
    \centering
    \includegraphics[width=0.45\textwidth]{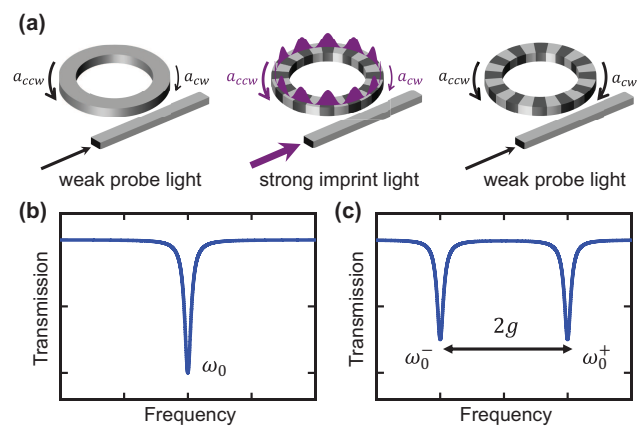}
    \caption{Optical mode splitting induced by photorefractive effect. (a) When strong light is launched to excite a LN ring cavity, an index pattern is imprinted in the microring by the standing-wave component of light intensity due to PR effect. The periodic index pattern leads to an enhancement of back scattering in the cavity. (b,c) Simulated spectrum of a resonance before (b) and after (c) index pattern generation. A mode splitting of $2g$ is generated due to strong coupling between CW mode ($a_{\rm{cw}}$) and CCW mode ($a_{\rm{ccw}}$), where $g$ is the mode coupling strength. }
    \label{fig_schematic}
\end{figure}

In this letter, we describe strong Bragg scattering induced by the PR effect in high-Q LN ring resonators measured at \SI{1.8}{\kelvin}. Cryogenic operation of LN microring resonators is critical for the exploitation of the strong Pockels nonlinearity of LN for microwave-to-optical photon conversions\cite{soltani2017efficient,mckenna2020cryogenic,shao2020integrated} and cryogenic-to-room-temperature data links\cite{youssefi2020cryogenic}. As the temperature decreases, the relaxation time of the PR effect increases from tens of milliseconds at room temperature \cite{jiang2017fast,sun2017nonlinear} to several days  at \SI{1.8}{\kelvin}\cite{von1978intrinsic}. Thus the PR effect induced electric field can semi-permanently modulate the refractive index of ring cavity. In particular, a periodic refractive index pattern similar to a Bragg-grating reflector can be built up when launching strong optical standing wave into selected modes of the cavity. Subsequently, when probed with a weak light, the microring exhibits mode splitting at phase matched wavelengths.
An illustration of such a mechanism is shown in Fig.\,\ref{fig_schematic}. A non-universal mode splitting of cavity resonances has been reported in optical resonator with fine lithographically engineered cavity \cite{lu2014selective}. Here we achieved selective mode splitting with all-optical control. Moreover, by strongly exciting other cavity modes, the imprinted index pattern could be redistributed, thus reconfigure the mode splitting to other wavelengths. These interesting observations suggest potential exploitation of PR effect for on-chip all-optically controlled photonic components. 

\begin{figure}[!t]
    \centering
    \includegraphics[width=0.45\textwidth]{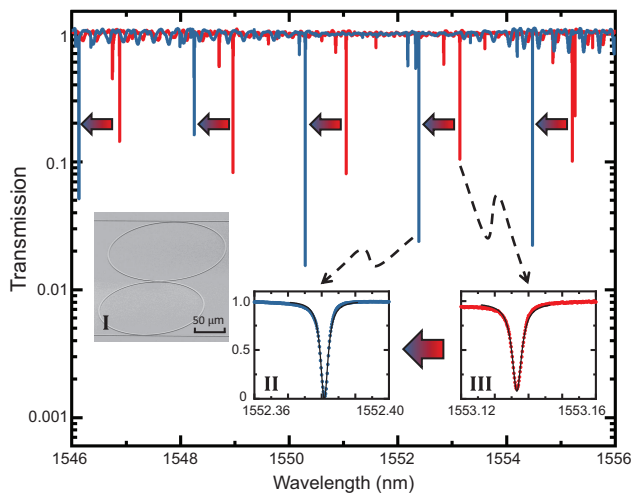}
    \caption{ Long-lived blue resonance shifts occur after periodic laser scanning across the transmission window of the grating couplers. Inset I shows an SEM image of the resonator. Inset II and III shows the zoomed-in spectrum of a TE00 mode after and before the periodic scanning, respectively.} 
    \label{fig_spectra}
\end{figure}

The device used in this work is a high-Q LN coupled double-ring  resonators (inset I of Fig.\,\ref{fig_spectra}) fabricated from a \SI{600}{\nano\meter} x-cut thin film LNOI wafer (from NANOLN), originally designed for microwave-to-optics conversion \cite{soltani2017efficient,mckenna2020cryogenic,shao2020integrated}. The coupled optical ring resonators have a width of \SI{1.6}{\micro\meter}, with radius \SI{80}{\micro\meter} and \SI{90}{\micro\meter} respectively,  and a coupling gap of \SI{0.7}{\micro\meter}. The device is patterned by electron beam lithography, with \SI{350}{\nano\meter}-thick LN etched through $\mathrm{Ar^+}$-based reactive ion etching. Detailed fabrication process can be found in our previous work \cite{lu2019octave}. In the final step, the device is coated with \SI{1.5}{\micro\metre} silicon dioxide using plasma enhanced chemical vapor deposition. 

In this double-ring device, 
when the wavelengths of the modes in two rings are close enough, the resonators are strongly coupled and support symmetric and anti-symmetric supermodes \cite{soltani2017efficient}. With optical power distributed in both rings, the two supermodes could be probed by a waveguide coupled to one of the rings. 
Although the specific device we employ here is more complex than a single microring, the impact of photorefractive effect and the underlying dynamics we elucidate here should apply to a range of other device geometries including simpler ring, racetrack or disk resonators. 

The chip is mounted on a set of attocube stages inside a closed-cycle cryostat and cooled down to \SI{1.8}{\kelvin}. The light output from a tunable laser diode (Santec-710) is sent to a variable optical attenuator (VOA) and followed by a polarization controller, then launched into the fridge via a standard single mode fiber. The light is coupled in and out from the on-chip waveguides through a pair of grating couplers designed to transmit TE polarized light. The light in the \SI{0.8}{\micro\meter} waveguide is coupled to the ring of \SI{80}{\micro\meter} radius with a coupling gap of \SI{1.0}{\micro\meter}. The insertion loss of the chip as well as the fiber inside fridge is \SI{23}{\decibel} and the output light is detected by a fiber-optic receiver.


The PR effect in LN is an action of several cascaded processes that induce a refractive index variation of the material in presence of light illumination \cite{hall1985photorefractive}. Due to the broken inversion symmetry of its crystal structure, when the LN is illuminated, a photocurrent is generated along crystalline z-direction through bulk photovoltaic effect. The migrated photo-induced charges are then trapped in defects of crystal, building up a space-charge field opposite to the direction of photocurrent. The electric field subsequently modulates the refractive index of the LN crystal through the Pockels effect. Specific to fabricated LNOI microcavities discussed here, the drop of refractive index results in a blue shift of resonance frequency with a relaxation time ranging from tens of milliseconds to several seconds at room temperature \cite{jiang2017fast,sun2017nonlinear}.

The relaxation time of the PR effect rises sharply when the device is cooled down to \SI{1.8}{\kelvin} \cite{von1978intrinsic,hall1985photorefractive}. As the cavity being illuminated at cryogenic temperature, the space-charge field induced by the PR effect accumulates and leads to long-lived modulation of resonances in the cavity. 
A continuous blue shift of all resonances is observed when we repeatedly scan the laser from \SI{1500}{\nano\meter} to \SI{1600}{\nano\meter} within the transmission window of the grating couplers with an input power of \SI{5}{\dBm} into to fridge (-6.5dBm on the input waveguide). 
The blue shift of resonance reaches a saturation after around thirty minutes of scanning. The normalized spectrum before scanning and after saturation is shown in Fig.\,\ref{fig_spectra}. The high-extinction mode group we study in this letter is the fundamental TE00 mode of the  ring resonator with radius of \SI{80}{\micro\meter}, which has an free spectral range (FSR) of ~\SI{2.1}{\nano\meter} around \SI{1550}{\nano\meter} and average loaded Q of $6\times 10^5$. A universal blue shift of \SI{0.76}{\nano\meter} is observed in the TE00 mode group. The saturation behaviour can be explained by the limited defect density available for charge excitation and trapping process in LN crystal \cite{hall1985photorefractive,sun2017nonlinear}. The average loaded Q of TE00 mode group remains approximately unchanged before and after the blue shift. We also record the time-dependent relaxation of resonances at \SI{1.8}{\kelvin}. The TE00 mode group experience a relaxation of \SI{0.21}{\nano\meter} after four days, and a characteristic time constant of more than ten days is obtained from the exponential fitting of the recorded shift trace. 

\begin{figure}[!t]
    \centering
    \includegraphics[width=0.45\textwidth]{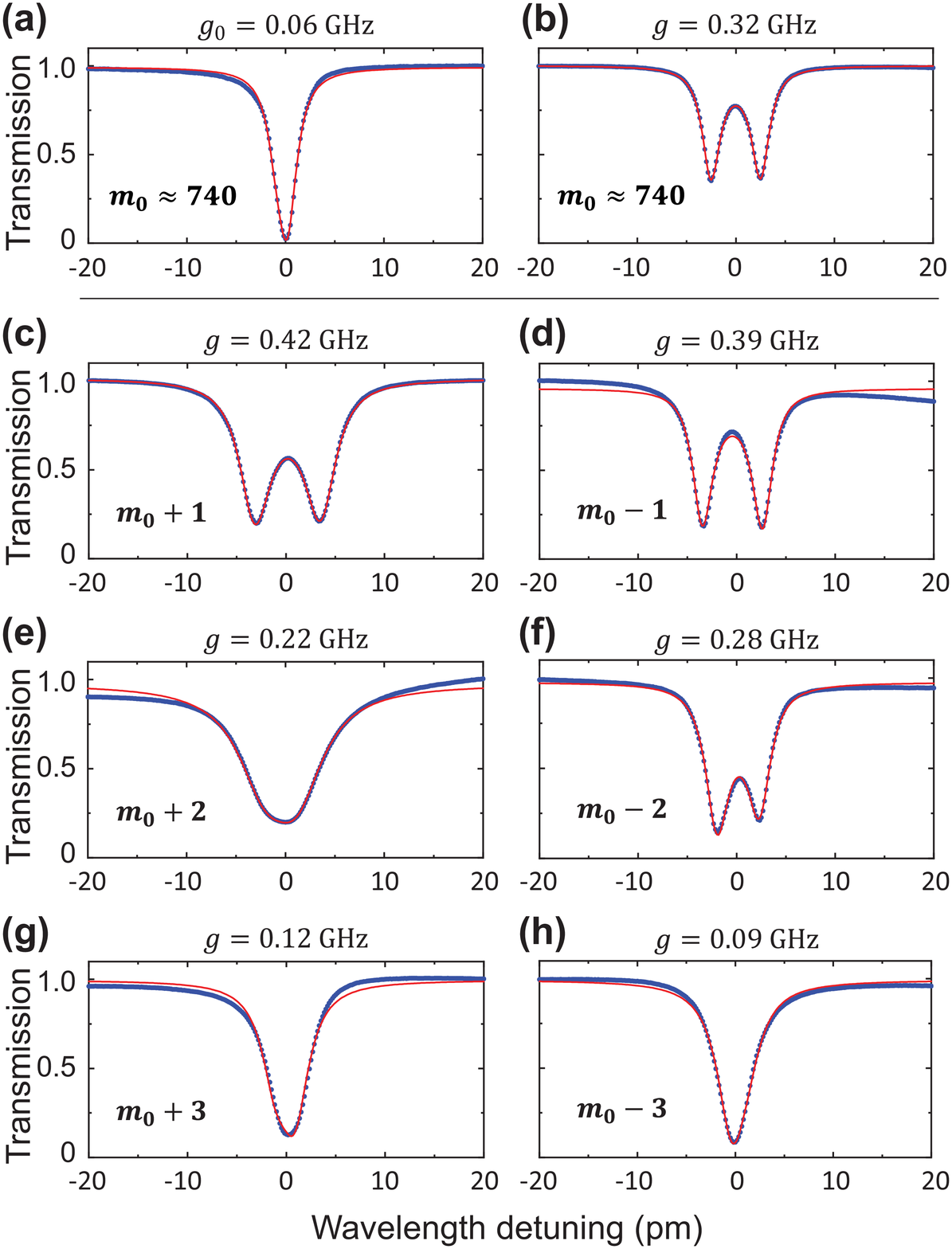}
    \caption{PR-induced mode splitting measured after strong illumination of the selected TE00 mode at \SI{1552.4}{\nano\meter}, with azimuthal mode number of $m_0$\,$\approx$740 in the \SI{80}{\micro\meter} radius ring. Transmission spectrum of mode $m_0$ before (a) and after (b) strong light illumination. 
    The fitted $g_0$ in (a) suggests a weak intrinsic back scattering in the ring resonator.
    (c) to (h) show the transmission spectra of the neighboring modes with azimuthal mode number $m_0 \pm 1$, $m_0 \pm 2$ and $m_0 \pm 3$ after the strong light illumination, respectively. The fitted back-scattering strengths $g$ is indicated in each figure. 
    }
    \label{fig_mode_number}
\end{figure}

After the saturation of universal blue shift induced by repeated scanning, the TE00 mode at \SI{1552.4}{\nano\meter} remains a lorentzian shape when probed with a weak light (\SI{-25}{\dBm} to the fridge), as shown in Fig.\,\ref{fig_mode_number}(a). Interestingly, after a strong light (\SI{5}{\dBm} to the fridge) is launched into this mode, the resonance experiences a frequency shift and the laser would be out of resonance in a few seconds. A mode splitting of the same mode could be observed in subsequent weak laser scan measurements [Fig.\,\ref{fig_mode_number}(b)]. This mode splitting of several resonance linewidths preserves for days in the absence of further strong light illumination, indicating its origin from the long-lived space charge field induced by PR effect.  This mode splitting of single resonance is attributed to the back scattering between the clockwise (CW) and counter clockwise (CCW) modes, which are renormalized into two standing-wave modes of different frequencies $\omega_0 \pm g$. Here $\omega_0$ is the original frequency of the resonance mode $m_0$ and $g$ describes the coupling strength between CW and CCW mode. In microring/microdisk cavity, this back scattering is normally contributed by the roughness of the cavity surface \cite{iwatsuki1984effect}. 
We found that the PR effect induced mode splitting is selective and not universal. It is only resolved in the TE00 modes with azimuthal mode number close to the imprinted mode. The imprinted TE00 mode at \SI{1552.4}{\nano\meter} has an azimuthal mode number of $m_0\approx740$. As shown in Fig.\,\ref{fig_mode_number}, after the imprint process, the largest mode splitting occurs in mode $m_0$ and $m_0 \pm 1$. As the azimuthal mode number difference increases, the mode splitting becomes weaker, and indistinguishable when the modes are more then 3\,FSRs away from the imprinted mode. 

 
In the following, we present a theoretical description of the mode splitting process.  Prior to the strong illumination of the selected cavity mode, an intrinsic weak Rayleigh back scattering already exists in the ring cavity due to surface roughness and inevitable fabrication imperfections  \cite{iwatsuki1984effect,borselli2005beyond}. 
When a strong light is launched in the microring, both CW and CCW traveling-wave modes are excited due to this intrinsic Rayleigh scattering. The cavity field can be expressed as
\begin{equation}
A(r,z,\phi)=A_{\rm{ccw}}(r,z) \Phi_{m_0}(\phi)+A_{\rm{cw}}(r,z) \Phi_{m_0}(-\phi)
\label{Eq_field}
\end{equation}

where $A_{\rm{cw}}$ and $A_{\rm{ccw}}$ represent the amplitude of CW and CCW imprint mode initially set by the launched and back-scattered light; $m_0$ is the azimuthal mode number of the imprint mode. Due to broken rotation symmetry in x-cut LN, $\Phi_{m}(\phi) \simeq \mathrm{exp}(i(m\phi-f\mathrm{sin}(2\phi)))$, where $f \simeq \pi R (n_o-n_e)/2 \lambda$ is determined by ring radius $R$ \cite{furst2016whispering}. For simplification, we set the initial phase to be 0 without loss of generality in the derivation. The optical field intensity $I$ is proportional to $|A|^2$, and therefore introducing a $\phi$-dependent standing-wave term, $I_{\rm{sw}}(\phi) \propto 2 |A_{\rm{ccw}} A_{\rm{cw}}| 
\Re[\Phi_{m_0}^2(\phi)]$.

We assume the photovoltaic effect will build up a space-charge field in proportion to the imprint light intensity along crystalline z-direction of \cite{sun2017nonlinear}, thus the $\phi$-dependent space-charge field on the cavity should satisfy $E_{sc}(\phi)=E_{sc,0} \Re[\Phi_{m_0}^2(\phi)]$. Considering the crystalline direction of x-cut LN film, the $\phi$-dependent refractive index modulation that inducing non-universal mode splitting can be calculated
\begin{multline}
\delta n(\phi)  \approx  \frac{E_{sc,0}}{4n_{\rm{eff}}} \Re[ (r_{33}+r_{13}) \Phi_{m_0}^2(\phi) + \\ \frac{(r_{33}-r_{13})}{2}(\Phi_{m_0-1}^2(\phi)+\Phi_{m_0+1}^2(\phi))]
\label{Eq_dn}
\end{multline}
\noindent where $n_{\rm{eff}}=[2 n_{o}^{2} n_{e}^{2}/(n_{o}^{2}+n_{e}^{2})]^{1/2}$ is the effective index of LN ring cavity, $r_{13}$ and $r_{33}$ are the electro-optic coefficient of LN. The PR-induced back scattering strength $g$ of $m$-th azimuthal mode is given by the integration of CW-CCW modal overlap weighted by the index modulation pattern, and the result can be simplified to 
\begin{equation}
g=\frac{\omega}{2n\pi}\int{\delta n(\phi)\Phi_{m_0}^2(\phi)d\phi}
\label{Eq_g}
\end{equation}
where $\omega$ is the resonance frequency. The value of $g$ could be derived  with Equation (\ref{Eq_dn}) and (\ref{Eq_g}),
\begin{equation}
g=\left\{
\begin{array}{crl}
\frac{(r_{33}+r_{31}) \omega E_{sc,0}}{8 n_{\rm{eff}}^2}     &      & m=m_0\\
\frac{(r_{33}-r_{31}) \omega E_{sc,0}}{16 n_{\rm{eff}}^2}     &      & m=m_0 \pm 1\\
0     &      & m \neq m_0, m_0 \pm 1
\end{array} \right. 
\end{equation}
which theoretically predict non-zero back scattering strength $g$ for modes $m=m_0$ and $m=m_0 \pm 1$, and the mode splitting value of mode $m_0$ should be higher than the value of mode $m_0 \pm 1$. In the experiment, the mode splitting of mode $m_0 \pm 1$ is comparable to mode $m_0$, and also splitting for mode $m_0 \pm 2$ is observed. This deviation between theoretical prediction and experimental data suggests that the imprinted pattern is not perfected matched with our prediction. The intrinsic roughness of the probed cavity mode, which is ignored in Eqs.\,(\ref{Eq_dn}), could also contribute to the backs-scattering process. The presence of bus waveguide and the second ring also breaks the rotational symmetry, leading to imperfections with Fourier components having 180$^o$ and 90$^o$ periodicity. Furthermore, our simple model only linearly relates the index pattern to the local optical intensity. Although it captures some of the features of the PR-induced mode splitting, further investigation is needed to understand the detailed space-charge generation and photon-trap coupling dynamics. A full-fledged model should also consider the microscopic spatial charge distribution in both longitudinal and transverse to the waveguides and self-consistently solve for space-field and the corresponding index modulation. We anticipate a z-cut single ring resonator will provide better symmetry and exhibit more distinctive mode-splitting features.


\begin{figure}[!t]
    \centering
    \includegraphics[width=0.45\textwidth]{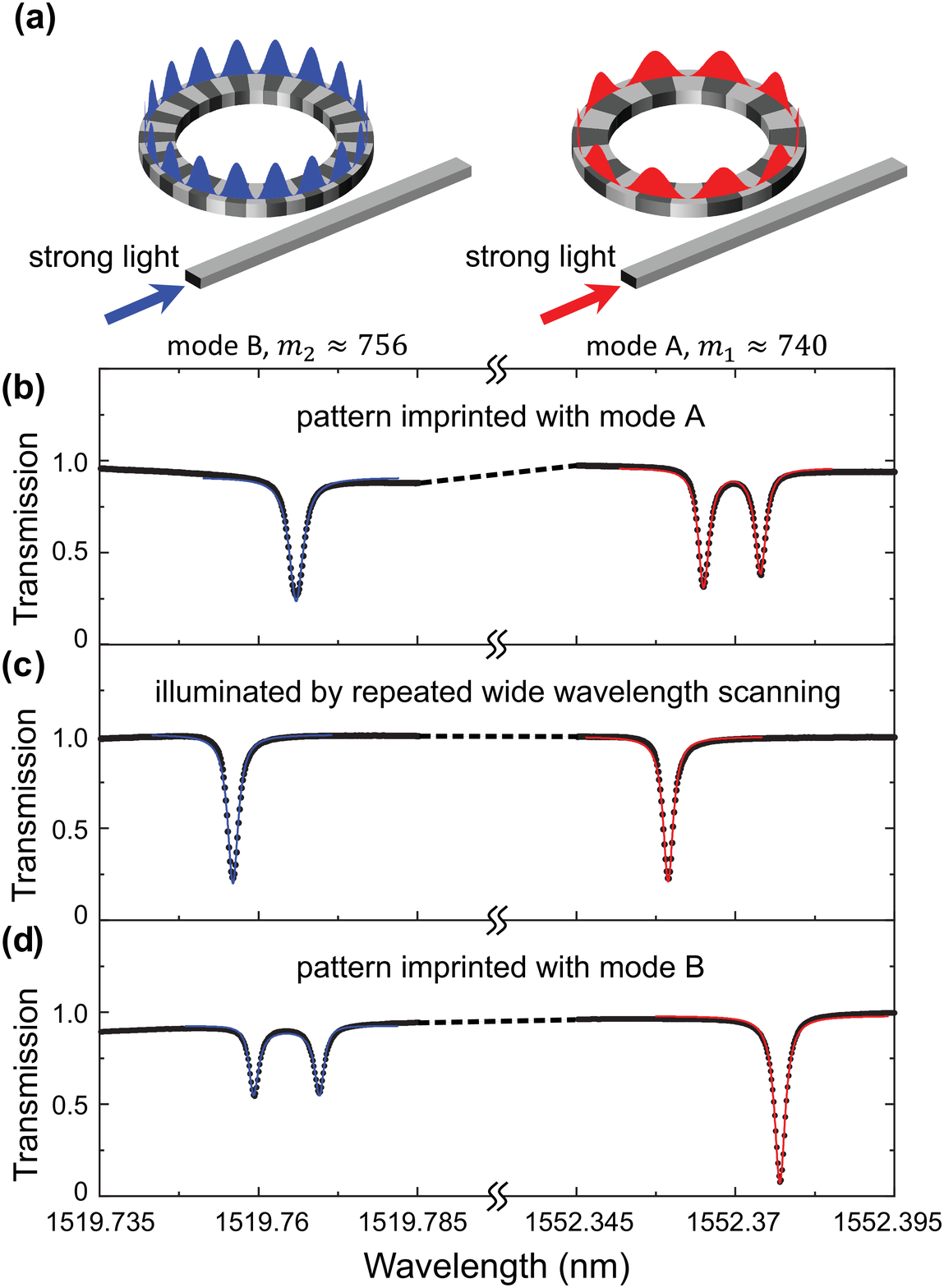}
    \caption{ (a) Schematic illustration of space-charge redistribution by strongly exciting different azimuthal modes. (b) to (d) display the transmission spectrum measured by a weak probe light during the reconfiguration sequence. (b) Mode splitting around of TE00
    mode A at \SI{1552.4}{\nano\meter} is first generated by launching strong light into mode A. (c) Mode splitting vanishes after the device is uniformly illuminated by repeated broadband laser scanning from \SI{1500}{\nano\meter} to \SI{1600}{\nano\meter}. (d) A strong light is re-launched into mode B at \SI{1519.7}{\nano\meter}, reconfiguring the mode splitting to mode B.
    }
    \label{fig_reconfig}
\end{figure}

The imprinted index pattern can be redistributed by shifting the strong excitation to other modes, thus selectively induce splitting in other modes at different wavelengths. A demonstration of the such reconfiguration process is shown in Fig.\,\ref{fig_reconfig}. An initial index pattern is imprinted by launching strong light (\SI{5}{\dBm} to fridge) into the mode A at \SI{1552.4}{\nano\meter}, introducing a mode splitting on this mode. By periodically scanning the laser wavelength from \SI{1500}{\nano\meter} to \SI{1600}{\nano\meter} covering 50\,FSR with an input power of \SI{5}{\dBm} for several times, the space-charge is smoothed out and mode splitting vanishes. A strong pump light is then launched into mode B at \SI{1519.76}{\nano\meter}, 16\,FSR away from mode A. As shown in Fig.\,\ref{fig_reconfig}(d), the mode splitting is then reconfigured to mode B, while the original mode A around \SI{1552.4}{\nano\meter} remains unchanged. Here we demonstrate that a the information of mode splitting could be written and erased through strong imprint light and read by weak probe light. This reconfigurable feature of PR effect at low temperature shows a potential which could be utilized as optically addressable switching or reconfigurable photonic components.

In conclusion, due to the extremely long relaxation time of the PR-induced space charges at cryogenic temperatures, the PR effect presents a significant challenge in nonlinear LN photonic devices that require strong optical pumping such as quantum frequency converters and microwave to optical converters. With active manipulation of the pump light, the space charges can be redistributed at much faster time scale for active manipulation of target optical modes. In particular, we show that by intensively illuminating the device, a periodic refractive index grating can be configured and readout with a subsequent weak probe light.  The device we characterized therefore possesses memory and reconfiguration capabilities that could be potentially utilized for on-chip all-optical reconfigurable photonic components.

~\\
\noindent\textbf{Funding.} This work is funded by ARO under grant number W911NF-18-1-0020. HXT acknowledges partial supports from NSF (EFMA-1640959), ARO (W911NF-19-2-0115) and the Packard Foundation. Funding for substrate materials used in this research was provided by DOE/BES grant under award number DE-SC0019406. 

~\\
\noindent\textbf{Acknowledgements.} The authors thanks Michael Rooks, Yong Sun, Sean Rinehart and Kelly Woods for support in the cleanroom and assistance in device fabrication. 

~\\
\noindent\textbf{Disclosures.} The authors declare no conflicts of interest.

\bibliography{reference}

\bibliographyfullrefs{reference}

\end{document}